\documentclass[twocolumn,aps,prl,superscriptaddress,graphicx, jmp,amsmath,amssymb,reprint]{revtex4-1} 

\usepackage{graphicx}
\usepackage[tight]{subfigure}
\usepackage{dcolumn}
\usepackage{bm}
\usepackage[mathlines]{lineno}
\usepackage{upgreek}

\newcommand{\BABARPubYear}     {12}
\newcommand{\BABARPubNumber}  {007}
\newcommand{\SLACPubNumber}   {15292}
\newcommand{\LANLNumber}      {1211.6271}

\input babarsym.tex

\def\BaBar{\babar\xspace}

\begin{document}

\preprint{\BaBar-PUB-\BABARPubYear/\BABARPubNumber}
\preprint{SLAC-PUB-\SLACPubNumber}

\begin{flushleft}
  \BaBar-PUB-\BABARPubYear/\BABARPubNumber \\
  SLAC-PUB-\SLACPubNumber \\
  hep-ex/\LANLNumber
\end{flushleft}

\title{Study of the reaction $e^+e^-\to\psi(2S)\pipi$ via initial-state radiation at \babar} 

%
\author{J.~P.~Lees}
\author{V.~Poireau}
\author{V.~Tisserand}
\affiliation{Laboratoire d'Annecy-le-Vieux de Physique des Particules (LAPP), Universit\'e de Savoie, CNRS/IN2P3,  F-74941 Annecy-Le-Vieux, France}
\author{J.~Garra~Tico}
\author{E.~Grauges}
\affiliation{Universitat de Barcelona, Facultat de Fisica, Departament ECM, E-08028 Barcelona, Spain }
\author{A.~Palano$^{ab}$ }
\affiliation{INFN Sezione di Bari$^{a}$; Dipartimento di Fisica, Universit\`a di Bari$^{b}$, I-70126 Bari, Italy }
\author{G.~Eigen}
\author{B.~Stugu}
\affiliation{University of Bergen, Institute of Physics, N-5007 Bergen, Norway }
\author{D.~N.~Brown}
\author{L.~T.~Kerth}
\author{Yu.~G.~Kolomensky}
\author{G.~Lynch}
\affiliation{Lawrence Berkeley National Laboratory and University of California, Berkeley, California 94720, USA }
% ...
\collaboration{The \babar\ Collaboration}
\noaffiliation

\begin{abstract}
We study the process $e^+e^-\to\psi(2S)\pipi$ with initial-state-radiation events produced at the PEP-II asymmetric-energy collider. The data were recorded with the \BaBar detector at center-of-mass energies at and near the $\Upsilon(\mathrm{nS})$ (n = 2, 3, 4) resonances and correspond to an integrated luminosity of  520\invfb. We investigate the $\psi(2S)\pipi$ mass distribution from 3.95 to 5.95 \gevcc, and measure the center-of-mass energy dependence of the associated  $e^+e^-\to \psi(2S)\pipi$ cross section. The mass distribution exhibits evidence of two resonant structures. A fit to the $\psi(2S)\pipi$ mass distribution corresponding to the decay mode $\psi(2S)\to J/\psi \pipi$ yields a mass value of $4340~\pm16$~(stat)~$\pm~9$~(syst)~\mevcc and a width of  $94~\pm~32$~(stat)~$\pm~13$~(syst)~\mev for the first resonance, and for the second a mass value of~$4669~\pm~21$~(stat)~$\pm~3$~(syst)~\mevcc and a width of  $104~\pm~48$~(stat)~$\pm~10$~(syst)~\mev. In addition, we show the $\pipi$ mass distributions for these resonant regions.\\

\pacs{}{\noindent \small {PACS numbers: 13.20.Gd, 13.25.Gv, 13.66.Bc, 14.40.Pq, 12.40.Yx, 12.39.Mk, 12.39.Pn, 12.39.Ki }}

\end{abstract}
\maketitle 


Many new $c\bar{c}$ or charmonium-like states have been discovered at the $B$-factories in the energy region above $D\overline{D}$ threshold. Of these, the $X(3872)$~\cite{ct:X3872}, $\chi_{c2}(2P)(3930)$~\cite{ct:Z3930}, $Y(3940)$~\cite{ct:Y3940}, and $Y(4260)$~\cite{ct:babar-Y} resonances are now well-established. Since the $Y(4260)$ is produced via initial-state radiation (ISR) in the reaction $e^+e^-\to \gamma_{\mathrm {ISR}} J/\psi \pipi$, it has  $J^{PC}=1^{--}$. In addition to the $Y(4260)$, two more $J^{PC}=1^{--}$ states, the $Y(4360)$ and the $Y(4660)$, have been reported in ISR production, via $e^{+}e^{-} \to \gamma_{\mathrm {ISR}}\psi(2S) \pi^+\pi^-$~\cite{Aubert:2006ge,:2007ea}. The $Y(4660)$ has been observed only in the Belle experiment~\cite{:2007ea}, and so it is important to confirm the existence of this state.\\
\indent In this paper we utilize the ISR mechanism to study the reaction $e^{+}e^{-} \to \psi(2S)\pipi$ in the center-of-mass (c.m.)\ energy (E$_{\mathrm{cm}}$) range 3.95 -- 5.95 \gev, where the $\psi(2S)$ decays to $J/\psi\pipi $ or to $ l^{+}l^{-}$, with $l^{+}l^{-}$ representing either 
$e^{+}e^{-}$ or $\mu^{+}\mu^{-}$.\\
\indent We use a data sample corresponding to an integrated luminosity of 520\invfb, recorded by the \babar~detector at the
SLAC PEP-II asymmetric-energy $e^+e^-$ collider operating at and near the c.m.~energies of the $\Upsilon(\mathrm{nS})$ (n = 2, 3, 4) resonances. The detector is described in detail elsewhere~\cite{ct:babar-detector}.  
Charged-particle momenta are measured in a tracking system consisting
of a five-layer, double-sided, silicon vertex-tracker (SVT) and a
40-layer central drift chamber (DCH), both coaxial with the 1.5--T 
magnetic field of a superconducting solenoid.
An internally reflecting ring-imaging Cherenkov detector, and specific ionization measurements from the SVT and DCH,
provide charged-particle identification (PID). A CsI(Tl) electromagnetic
calorimeter (EMC) detects and identifies photons and electrons. Muons are identified using information from the instrumented
flux-return system.\\
\begin{figure}[htbp!]
\centering
    \includegraphics[width=7.8cm,height=5cm]{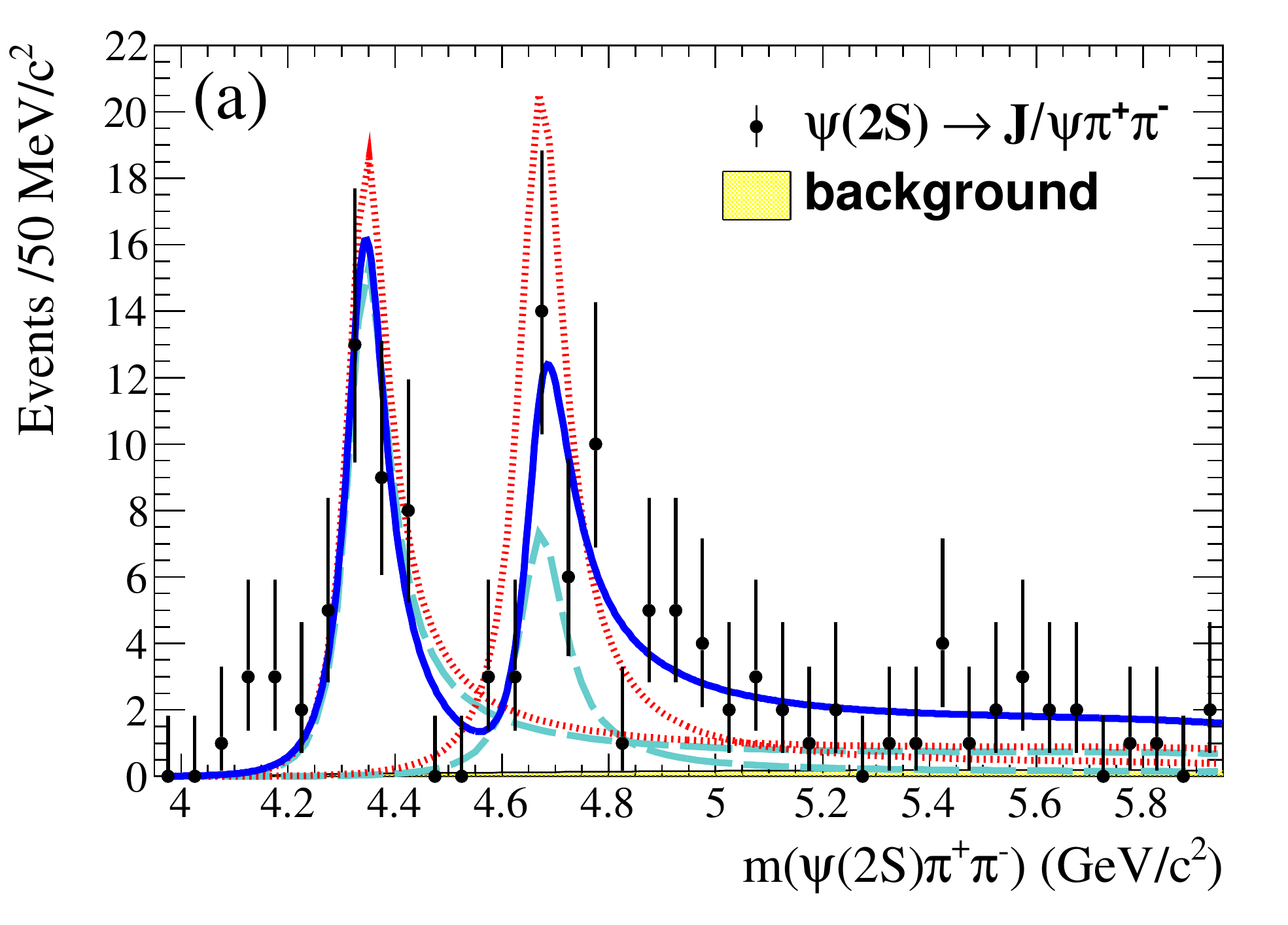}	
     \includegraphics[width=7.8cm,height=5cm]{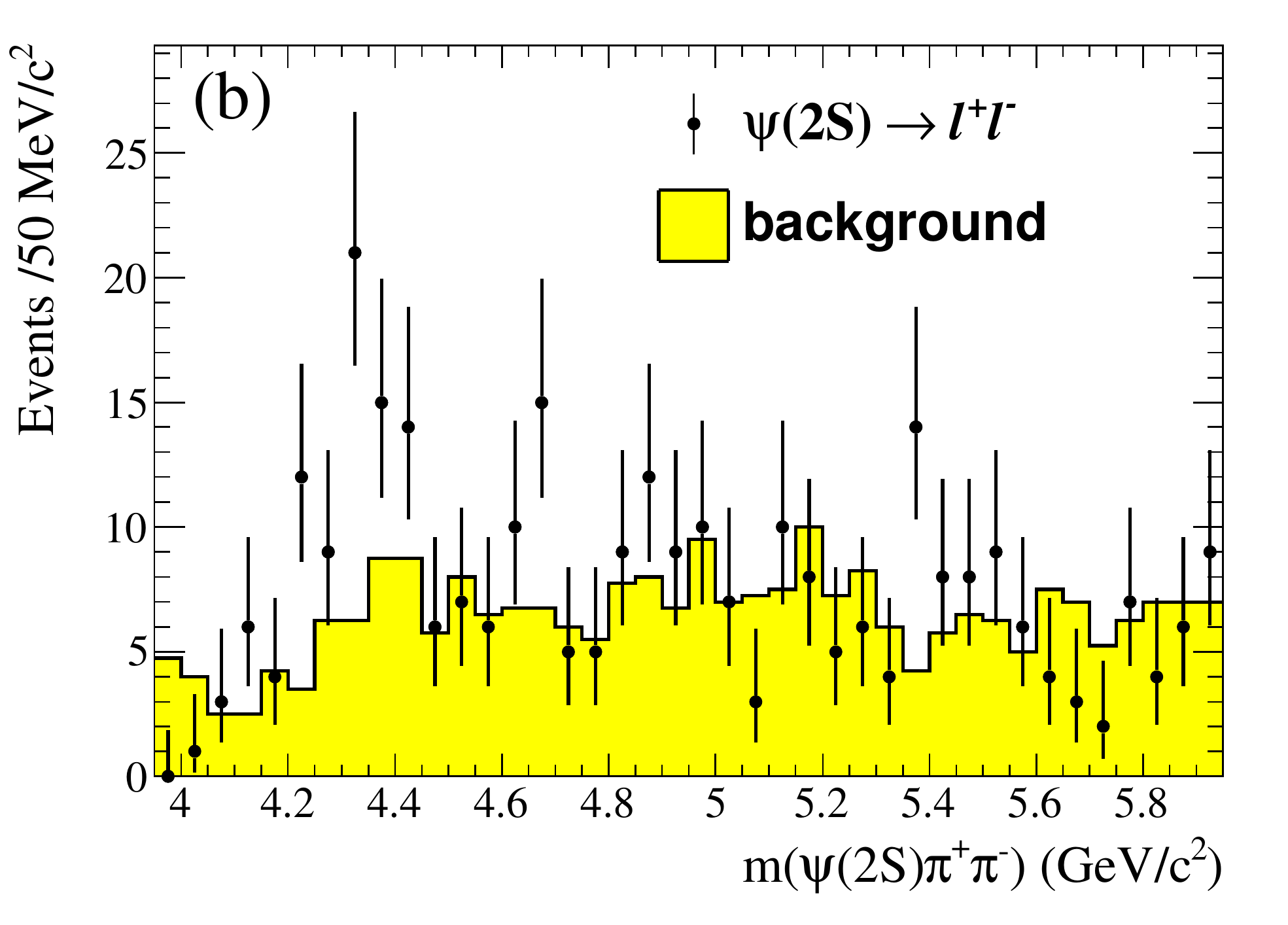}
     \includegraphics[width=7.8cm,height=5cm]{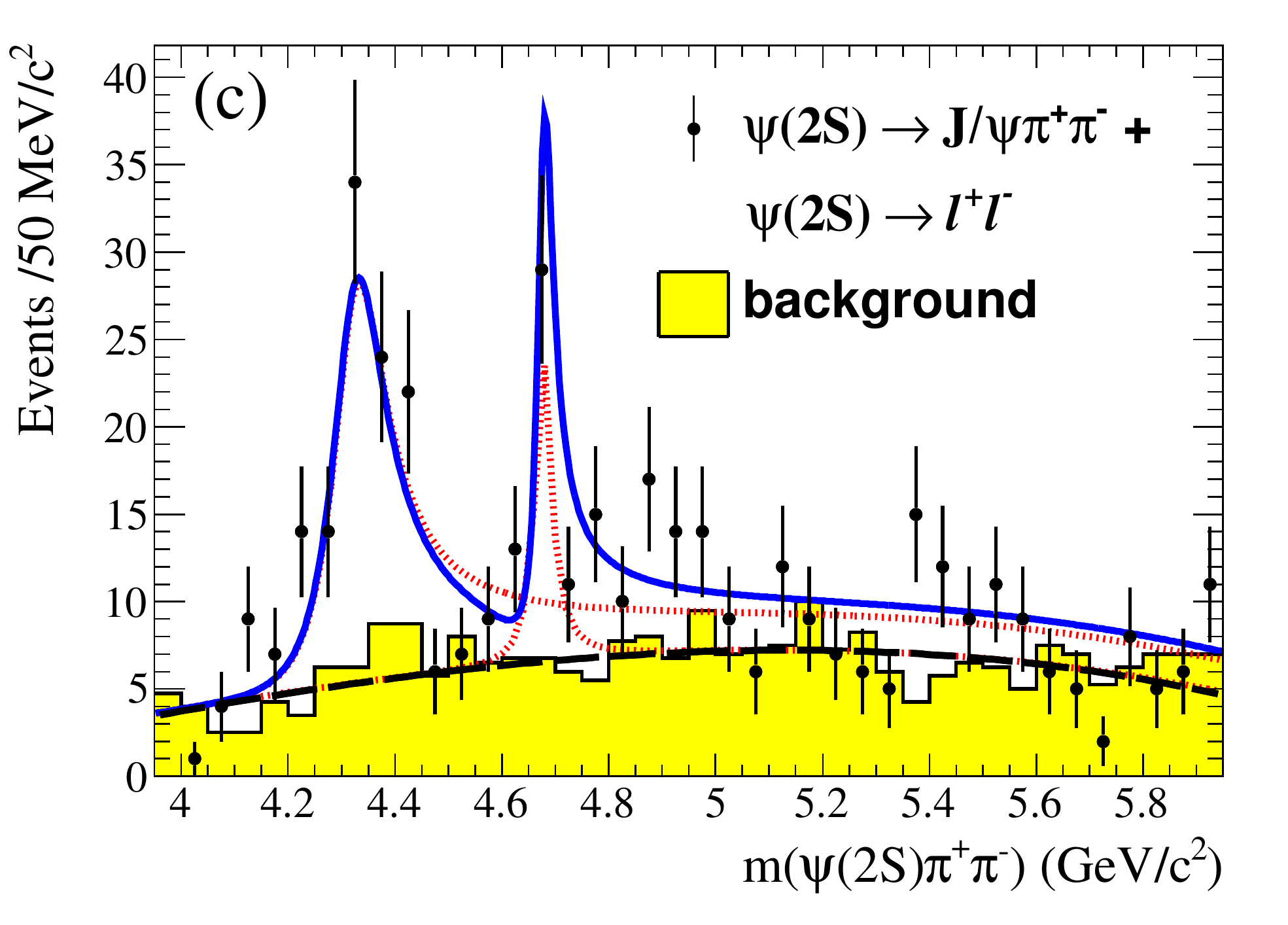}\\ 	
     \caption{(a) The $\psi(2S)\pipi$ invariant mass distribution from the kinematic threshold to  5.95 \gevcc for  $\psi(2S) \to J/\psi \pipi$; the points with error bars represent the data in the $\psi(2S)$ signal region, and the shaded histogram is the background estimated from the  $\psi(2S)$ sideband regions. The solid curve shows the result of the fit described in the text. The dashed (dotted) curves indicate the individual resonant contributions for constructive (destructive) interference. (b) The corresponding distributions for $\psi(2S) \to l^{+}l^{-}$. (c) The combined $\psi(2S)\pipi$ invariant mass distribution for  $\psi(2S) \to J/\psi \pipi$ and $\psi(2S)\to l^{+}l^{-}$. The solid curve shows the result of the fit. The dashed curve represents the background, while the dotted curves indicate the individual resonant contributions. There is only one solution in this case.}
  \label{fig:fit}
\end{figure}
\indent We reconstruct events corresponding to the reaction $e^+e^- \to \gamma_{\mathrm{ISR}} \psi(2S)\pipi$,
where $\gamma_{\mathrm{ISR}}$ represents a photon that is radiated from the 
initial-state $e^{\pm}$, thus lowering the c.m.\ energy of the $e^+e^-$ collision which produces the $ \psi(2S)\pipi$ system. 
We do not require observation of the ISR photon, since it would be detectable in the EMC for only $\sim$ 15\% of the events.\\
\indent For the $\psi(2S) \to J/ \psi\pipi$ decay mode, we select events containing exactly six charged-particle tracks, and reconstruct $\jpsi$ candidates via their decay to $e^+e^-$ or $\mu^+\mu^-$.
For each mode, at least one of the leptons  must be identified on the basis of PID information.
When possible, electron candidates are combined with photons to recover bremsstrahlung energy loss in order to improve the  $\jpsi$
momentum measurement.
An $e^+e^-$ pair with invariant mass within ($-$60,+45)\mevcc of the nominal $\jpsi$ mass~\cite{ct:PDG} is accepted as a $\jpsi$ candidate, as is a $\mu^+\mu^-$ pair with mass within ($-$45,+45)\mevcc of this value. Each $\jpsi$ candidate is subjected to a geometric fit in which the decay vertex is constrained to the $e^+e^-$ collision axis within the interaction region; the $\chi^{2}$-probability of the fit must be greater than $0.001$. An accepted $\jpsi$ candidate is kinematically constrained to the nominal $\jpsi$  mass~\cite{ct:PDG}, and combined with a pion pair to form a $J/\psi\pipi$ candidate. 
The $J/\psi\pipi$ combinations with invariant mass within 10 $\mevcc$ of the nominal $\psi(2S)$ mass are taken as $\psi(2S)$ candidates, and hereafter we refer to this as ``the $\psi(2S)$ signal region".
The $\psi(2S)$ candidate is refit requiring that the $\chi^{2}$-probability for the vertex fit be greater than $0.001$. It is then combined with two additional pions of opposite charge, each of which is identified using PID information, to reconstruct a $\psi(2S)\pipi$ candidate. A further geometric fit with the $\psi(2S)$ candidate mass-constrained to the nominal mass value ~\cite{ct:PDG} is performed. Candidates with $\chi^{2}$-fit probability greater than 0.001 are retained for further analysis.\\
\indent For the decay mode $\psi(2S)\to l^{+}l^{-}$, we select events containing exactly four charged-particle tracks, and reconstruct $\psi(2S)$ candidates via their decay to $e^+e^-$ or $\mu^+\mu^-$.
An $e^+e^-$ ($\mu^+\mu^-$) pair with invariant mass within ($-$40,+30)\mevcc (($-$30,+30)\mevcc) of the nominal $\psi(2S)$ mass~\cite{ct:PDG} is accepted as being within the $\psi(2S)$ signal region. Each such candidate is subjected to the same geometrical fit and mass constraint procedure as applied for the $\psi(2S)\to J/\psi\pipi$ mode. A surviving candidate is combined with a pion pair to form a $\psi(2S)\pipi$ candidate.
\begin{table}
\centering
\caption{Results of the fit to the $\psi(2S)\pipi$ invariant mass distributions for $\psi(2S) \to J/\psi\pipi$. The first errors are statistical and the second systematic; ${\cal B}\times \Gamma_{ee}$
is the product of the branching fraction to $\psi(2S)\pipi$ and the $e^{+}e^{-}$ partial width (in $eV$), and $\phi$ is the relative phase between the two resonances (in degrees).}
\label{tabfit}
\begin{tabular}{ccc}
  \hline
  \hline
  Parameters & First Solution & Second Solution\\
    & (constructive  & (destructive \\
     &  interference) & interference)\\	
\hline
Mass Y(4360) ($\mathrm{MeV/c^{2}}$)   &\multicolumn{2}{c}{4340 $\pm$ 16 $\pm$ 9 }    \\
Width Y(4360) ($\mathrm{MeV}$) & \multicolumn{2}{c}{94 $\pm$ 32  $\pm$ 13 }    \\
${\cal B}\times \Gamma_{ee}(Y(4360))~(\mathrm{eV})$& 6.0 $\pm$ 1.0 $\pm$ 0.5  &~~ 7.2 $\pm$ 1.0 $\pm$ 0.6   \\
Mass Y(4660) ($\mathrm{MeV/c^{2}}$) & \multicolumn{2}{c}{ 4669 $\pm$ 21 $\pm$ 3 }  \\
Width Y(4660) ($\mathrm{MeV}$) &  \multicolumn{2}{c}{ 104 $\pm$ 48 $\pm$ 10  }   \\
${\cal B}\times \Gamma_{ee}(Y(4660)) ~(\mathrm {eV})$ & 2.7 $\pm$ 1.3  $\pm$ 0.5  & ~~7.5 $\pm$ 1.7  $\pm$ 0.7 \\
$ \phi ~(^{\circ})$  & 12 $\pm$ 27 $\pm$ 4   &-78 $\pm$ 12 $\pm$ 3\\
\hline
\hline
\end{tabular}
\end{table}
\begin{table}
\centering
\caption{Results of the fit to the combined $\psi(2S)\pipi$ invariant mass distributions for $\psi(2S) \to J/\psi\pipi$ and $\psi(2S) \to l^{+}l^{-}$. The first errors are statistical and the second systematic; ${\cal B}\times \Gamma_{ee}$ is the product of the branching fraction to $\psi(2S)\pipi$ and the $e^{+}e^{-}$ partial width (in $eV$), and $\phi$ is the relative phase between the two resonances (in degrees).}
\label{tabfitll}
\begin{tabular}{ccc}
  \hline
  \hline
  Parameters & ~~~ & Solution \\
\hline
Mass Y(4360) ($\mathrm{MeV/c^{2}}$) &~~~ &4318$^{+15}_{-19}\pm$ 3    \\
Width Y(4360) ($\mathrm{MeV}$)& ~~~&123 $\pm$ 20  $\pm$ 13   \\
${\cal B}\times \Gamma_{ee}(Y(4360))~(\mathrm {eV})$ & ~~~& 7.4 $\pm$ 0.9 $\pm$ 0.7    \\
Mass Y(4660)($\mathrm{MeV/c^{2}}$) &~~~& 4667$^{+6}_{-7}\pm 2 $  \\
Width Y(4660) ($\mathrm{MeV}$) &~~~&   $36^{+32}_{-14}$  $\pm 4$ \\
${\cal B}\times \Gamma_{ee}(Y(4660)) ~ (\mathrm {eV})$ &~~~&1.4 $\pm$ 0.5  $\pm$ 0.2   \\
$ \phi ~(^{\circ})$ & ~~~& 25 $\pm$ 21 $\pm$ 2  \\
\hline
\hline
\end{tabular}
\end{table}
\\
\indent For $\psi(2S)\to J/\psi \pipi$~($\psi(2S)\to l^{+}l^{-}$),~the~difference~between~the~c.m.~momentum of the hadronic $\psi(2S)\pipi$ system and the value expected for an ISR event $(i.e.~(s-m^{2})/2\sqrt{s}$, where $m$ is the $\psi(2S)\pipi$ invariant mass) must be in the range ($-$0.10,+0.70)~$\gevc$ (($-$0.70,+0.60)~$\gevc$) to be consistent with an ISR photon. We require the transverse component of the missing momentum to be less than 2.0\gevc (1.7\gevc). If the ISR photon is detected in the EMC, its momentum vector is added to that of the $\psi(2S)\pipi$ system in calculating the missing momentum. For the events for which $\psi(2S)\to e^{+}e^{-}$, the candidate $\pipi$ system has a small contamination due to $e^{+}e^{-}$ pairs from photon conversions. We compute the pair invariant mass $m_{e^{+}e^{-}}$, with the electron mass assigned to each pion candidate, and remove candidates with $ m_{e^{+}e^{-}} <100\mevcc$.\\
\indent For events with multiple $\psi(2S)$ candidates, we select the combination that has candidate mass closest to the $\psi(2S)$ nominal mass value~\cite{ct:PDG}. We estimate the remaining background for the decay mode $\psi(2S) \to J/\psi \pipi$ using events that have a $J/\psi \pipi$ mass in either of the $\psi(2S)$ sideband regions (3.566,~3.666) or (3.706,~3.806) $\gevcc$. 
For the decay mode $\psi(2S)\to e^{+}e^{-}$, the corresponding regions are (3.476,~3.576) and (3.776,~3.876) $\gevcc$, while for $\psi(2S)\to \mu^{+}\mu^{-}$ the sideband regions are (3.516,~3.596) and (3.776,~3.856) $\gevcc$.\\
\indent Figure~\ref{fig:fit} shows the $\psi(2S)\pipi$ invariant mass distributions for the selected $\psi(2S)$ events corresponding to the decays (a)~$\psi(2S) \to \jpsi\pipi$, (b)~$\psi(2S) \to l^{+}l^{-}$, and (c)~ the combined sample for~$\psi(2S) \to \jpsi\pipi$ and~$\psi(2S) \to l^{+}l^{-}$. The background is estimated from the $\psi(2S)$ mass sidebands as described above. In Fig. \ref{fig:fit} two structures are evident, the first near 4.35 $\gevcc$, and the second near 4.65 $\gevcc$. We attribute these peaks to the $Y(4360)$~\cite{Aubert:2006ge} and to the $Y(4660)$~\cite{ :2007ea},~respectively. We first fit the distribution shown in Fig.~\ref{fig:fit}(a) in order to extract the parameter values of the resonances. We then perform a second fit to the combined $\jpsi\pipi$ and $l^{+}l^{-}$ data of Fig.~\ref{fig:fit}(c), where the signal yields are larger, but where these come at the cost of the large background associated with the dilepton channels. For both distributions we perform an unbinned, extended-maximum-likelihood fit to the $\psi(2S)\pipi$ mass distribution from the signal region, and simultaneously to the background mass distribution. We describe the latter by a fourth-order polynomial in $\psi(2S)\pipi$ mass, $m$, for the fit to the data of Fig.~\ref{fig:fit}(a), and by a third-order polynomial for the fit to the data shown in Fig.~\ref{fig:fit}(c). \\
\indent The mass dependence of the signal function is given by $ f(m) = \epsilon (m) \cdot{\cal L}(m)\cdot \sigma(m)$; $ \epsilon (m)$ is the mass-dependent signal-selection efficiency obtained from a MC simulation which uses a $\psi(2S)\pipi$ phase space distribution; its value increases from 1\% at 3.95 \gevcc to 12\% at 5.95 \gevcc for $\psi(2S) \to \jpsi\pipi$ and from 1\% at 3.95 \gevcc to 14\% at 5.95 \gevcc for~$\psi(2S) \to l^{+}l^{-}$. The function ${\cal L}(m)$ is the mass-distributed luminosity~\cite{Kuraev:1985hb} (we ignore the small corrections due to initial-state emission of additional soft photons);  ${\cal L}(m)$ increases from 102 pb$^{-1}$/50 MeV to 202 pb$^{-1}$/50 MeV from 3.95 \gevcc to 5.95 \gevcc.\\
\indent The cross section, $\sigma (m)$, is described by the following function, which takes into account the possibility of interference between the two resonant amplitudes, since they have the same quantum numbers ($J^{PC}=1^{--}$):
\begin{figure}[htbp]
\centering
    \includegraphics[width=7.8cm,height=5cm]{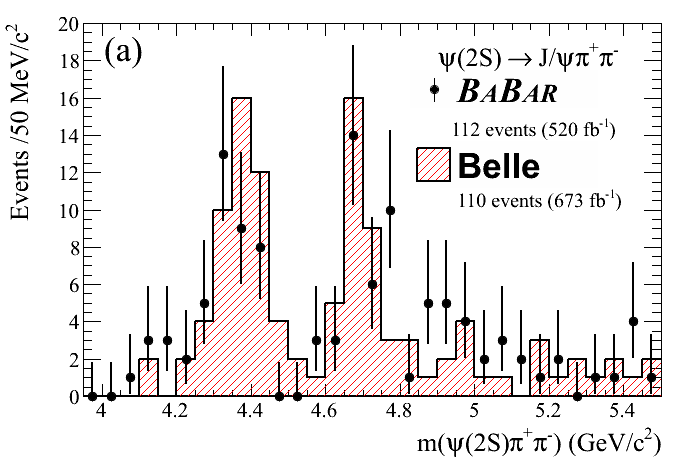}	
    \includegraphics[width=7.8cm,height=5cm]{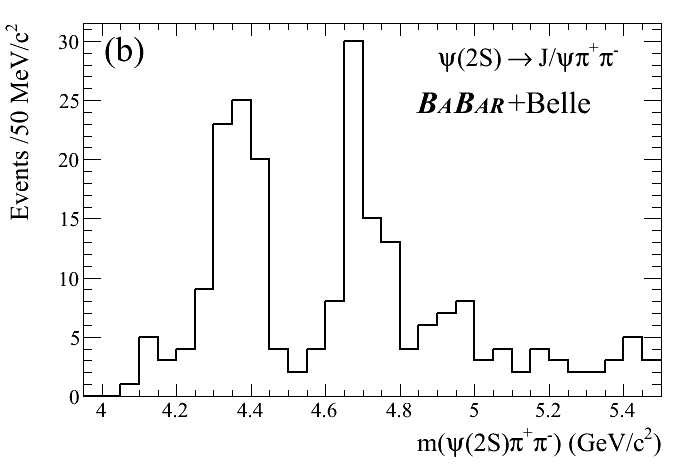}	\\
        \caption{(a) The comparison between the observed $ \psi(2S)\pipi$ ($\psi(2S)\to  J/\psi\pipi$) invariant mass spectrum from \babar~(dots) and that from Belle~(hatched histogram). (b) The combined ~\babar~and Belle~$\psi(2S)\pipi$ ($\psi(2S)\to  J/\psi\pipi$) invariant mass spectrum.  }
  \label{fig:comparison}
\end{figure}
\begin{figure}[htbp]
\centering
    \includegraphics[width=7.8cm,height=5cm]{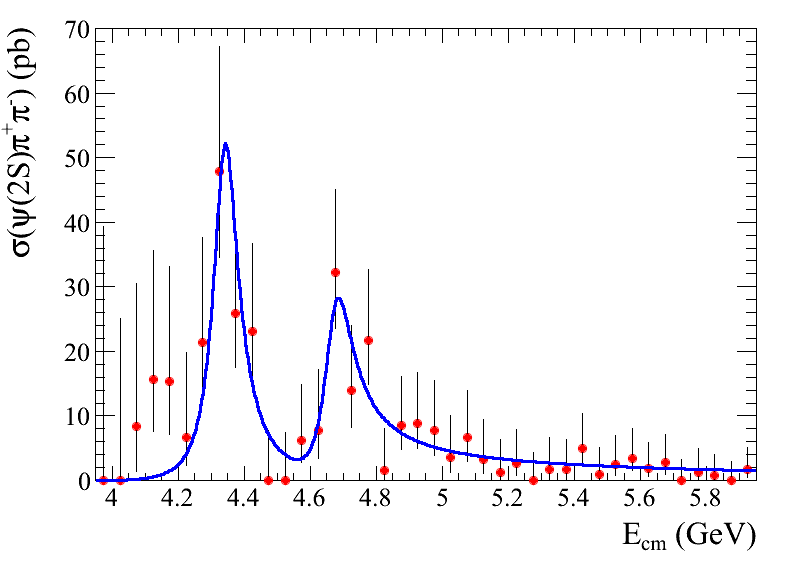}\\	
          \caption{The cross section for the reaction $e^{+}e^{-}\to \psi(2S)\pipi$ as a function of c.m. energy obtained by using Eq.~(\ref{eq:Cross}) (points with error bars); the curve shows the c.m. energy dependence which results from the fit to the data of Fig.~\ref{fig:fit}(a).  }
  \label{fig:crosssection}
\end{figure}
\begin{equation}
\label{pdf}
\begin{split}
\sigma(m)=& \frac{12\pi C}{m^{2}} \cdot |A_{1}(BW)\cdot \sqrt{\frac{PS(m)}{PS(m_{1})}} \\
              &+A_{2}(BW)\cdot \sqrt{\frac{PS(m)}{PS(m_{2})}}\cdot e^{i\phi}|^{2} \\
              \end{split}
\end{equation}
\noindent where $C=0.3894 \cdot 10^{9}~\mathrm{GeV^{2}~pb}$, and $PS(m)$ represents the mass dependence of $\psi(2S)\pipi$ phase space; $\phi$ is the relative phase between the amplitudes $A_{1}$ and $A_{2}$. The complex amplitude $A_{j}$ is written as 
\begin{equation}
A_{j}(BW)=\frac{m_{j}\sqrt{(\Gamma_{e^{+}e^{-}}\cdot \Gamma_{\psi(2S)\pi^{+}\pi^{-}})_{j}}}{m^{2}_{j}-m^{2}-im_{j}\Gamma_{j}}
\end{equation}
where $m_{j}$ is the resonance mass and $\Gamma_{j}$ its total width; $(\Gamma_{e^{+}e^{-}}\cdot \Gamma_{\psi(2S)\pi^{+}\pi^{-}})_{j}$ is the product of the partial widths to $e^{+}e^{-}$ and to $\psi(2S)\pi^{+}\pi^{-}$.\\
\indent In the fit procedure $f(m)$ is convolved with a Gaussian resolution function obtained from MC simulation. This function has root-mean-squared (r.m.s.)~deviation which increases linearly from 2\mevcc at $\sim$ 3.95 \gevcc  to 5\mevcc at $\sim$ 5.95  \gevcc. In the likelihood function, when the fit is performed to the $\psi(2S)\to J/\psi\pipi$ data, $\sigma(m)$ is multiplied  by ${\cal B}(\psi(2S)\to J/\psi \pi^{+}\pi^{-}) \times {\cal B}(J/\psi \to l^{+}\l^{-})$, since the fit is to the corresponding observed events. Similarly, for $\psi(2S)\to l^{+}l^{-}$, $\sigma(m)$ is multiplied by ${\cal B}(\psi(2S)\to \l^{+}\l^{-})$, where $l=e$ or $\mu$, in fitting the data. The results of the fits are shown in Fig.~\ref{fig:fit}(a) and in Fig.~\ref{fig:fit}(c), and the extracted parameters are summarized in Tables \ref{tabfit} and \ref{tabfitll}, respectively. The significance of the $Y(4660)$ signal for both fits is greater than 5$\sigma$ where $\sigma$ is the standard deviation. For the fit to the distribution in Fig.~\ref{fig:fit}(a), we obtain two solutions, one corresponding to constructive interference and one to destructive interference between the resonant amplitudes. The mass and the width values of the resonances are the same for each solution. However, the values of $\Gamma_{e^{+}e^{-}} \times {\cal B}(\psi(2S)\to J/\psi \pipi)$ and $\phi$ are different (see Table \ref{tabfit}), although the maximum likelihood value is exactly the same for each fit. For the fit to the distribution in Fig.~\ref{fig:fit}(c), only a solution showing constructive interference is obtained, for which the parameter values are consistent within error with those for the first solution in Table~\ref{tabfit}. The results summarized in Table~\ref{tabfit} agree well with those obtained in the Belle analysis~\cite{ :2007ea}, for which the data
 sample is about the same size as that for the $\psi(2S)\to J/\psi\pipi$  decay mode in the present analysis (see Fig.~\ref{fig:comparison}(a)).
  We infer that, even if our data sample for this mode were doubled in size, the ambiguity in the
 fit results would persist. The inclusion of the $\psi(2S)$ dilepton decay modes increases the signal sample by~40\% over that for the $\psi(2S)\to J/\psi\pipi$ mode alone. This increase is obtained at the cost of introducing a background contribution which is larger by~50\% than the combined signal
 sample. The fit to this sample yields only one solution (Table~\ref{tabfitll}). However, the comparison of our results for the $\psi(2S)\to J/\psi\pipi$
 analysis to those from the Belle analysis~\cite{ :2007ea} leads us to conclude that the apparent resolution of the fit ambiguity is due, not to the slightly
 increased signal sample, but rather to the presence of the large background shown in Fig.~\ref{fig:fit}(c). For this reason we discount the results
 summarized in Table \ref{tabfitll}, and confine our attention to the results from $\psi(2S)\to J/\psi\pipi$ decay in the remainder of the analysis.\\
\indent The fit results of Table \ref{tabfit} and the  $\psi(2S)\pipi$ invariant mass spectrum of Fig~\ref{fig:comparison}(a) agree very well with those obtained by the Belle Collaboration~\cite{ :2007ea}. Each distribution (Fig.~\ref{fig:comparison}(a)) shows evidence of two resonant signals (note that the Belle distribution ends at 5.5 \gevcc). This is even more apparent in Fig.~\ref{fig:comparison}(b), where we have added the distributions to obtain a mass spectrum corresponding to an integrated luminosity of $\sim$ 1.2 $\mathrm{ab^{-1}}$. The existence of two structures is quite clear, and there is even a hint of some activity in the vicinity of 5 \gevcc. \\
\indent For the decay mode $\psi(2S)\to J/\psi \pipi$, we calculate the $e^{+}e^{-} \rightarrow \psi(2S)\pi^{+}\pi^{-}$ cross section after background subtraction
for each $\psi(2S)\pi^{+}\pi^{-}$ mass interval, $i$,  using 
\begin{linenomath}
\begin{equation}\label{eq:Cross}
    \sigma_{i} = \frac{n_{i}^{\mathrm{obs}}-n_{i}^{\mathrm{bkg}}}{\epsilon_{i} \cdot {\cal L}_{i} \cdot  {\cal B}} ~,
\end{equation}
\end{linenomath}
\noindent 
where $n_{i}^{\mathrm{obs}}$ is~the number of observed events, $n_{i}^{\mathrm{bkg}}$ is the number of background events, $\epsilon_{i}$ is the average efficiency, and ${\cal L}_{i}$ the integrated luminosity~\cite{Kuraev:1985hb} for interval $i$; ${\cal B}$ represents the product ${\cal B} (\psi(2S) \rightarrow J/\psi \pi^{+}\pi^{-})\cdot {\cal B}  (J/\psi \rightarrow l^{+}l^{-})$. The resulting dependence of the cross section on c.m.~energy is shown in Fig.~\ref{fig:crosssection}. We sum over the data points in Fig.~\ref{fig:crosssection} and obtain a model-independent integrated cross section value of $311^{+76}_{-30}$ (stat) $\pm $11 (syst) pb for the region 3.95--5.95 GeV. The curve shown in Fig.~\ref{fig:crosssection} results from the fit to the data of Fig.~\ref{fig:fit}(a), and provides an adequate description of the measured cross section. 
\begin{figure}[!h]
\centering
      \includegraphics[width=7.8cm,height=5cm]{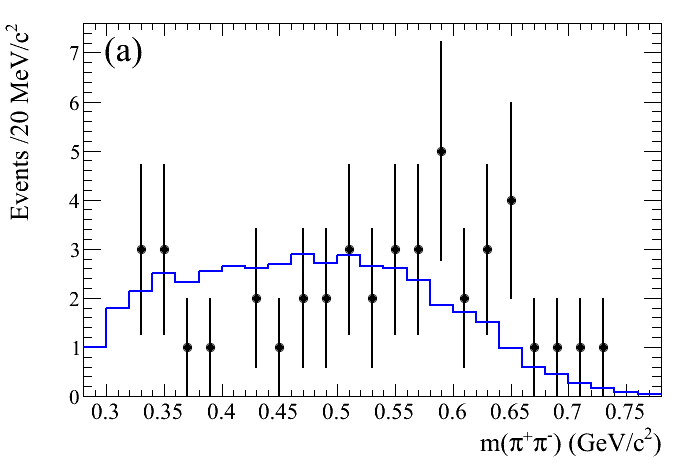}
     \includegraphics[width=7.8cm,height=5cm]{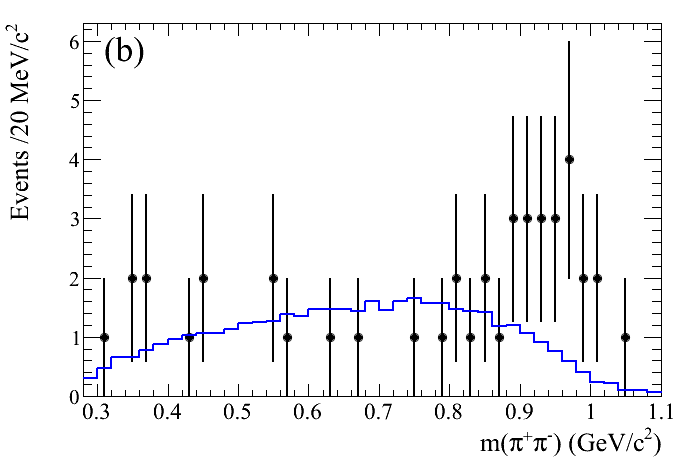}\\
     \caption{The \pipi invariant mass spectrum for the $\psi(2S)\to J/\psi\pipi$ channel in the $\psi(2S)\pipi$ mass region (a)  4.0--4.5 \gevcc, and
     (b) 4.5--4.9 \gevcc. The histogram represents a MC distribution corresponding to the decay according to phase space of (a) one resonance with a mass of 4.360 \gevcc and width 70 MeV, and (b) one resonance with a mass of 4.660 \gevcc and width 50 MeV. Each histogram is normalized to the corresponding data sample.}
  \label{fig:pipimass}
\end{figure}
\begin{table} 
\caption{\label{tab:systematic4360} Systematic uncertainty estimates for the parameters used in the fit to the data of Fig.~\ref{fig:fit}(a).} 
\begin{ruledtabular} 
\begin{tabular}{ccccc} 
Source & $\Gamma_{e^+e^-}\cdot{\cal B}$ & $\Gamma_{e^+e^-}\cdot{\cal B}$ & Mass  & $\Gamma$   \\ 
 &$(\%)$ & $(\%)$ & (\mevcc) &   (\mev)  \\ 
 & $_{\mathrm {(constructive}}$ & $_{\mathrm{(destructive}}$  &   &   \\ 
  &$ _{\mathrm{interference)}}$ & $ _{\mathrm{interference)}}$  &  &  \\ 
\hline 
Fit procedure & & &  &  \\ 
 for the Y(4360)& $\pm2.5$ &$\pm$ 1.4  & $\pm$9 & $\pm$ 13 \\ 
Fit procedure &  & & \\ 
for the Y(4660)& $\pm14$ &$\pm$ 3.3  & $\pm$3 & $\pm$ 10 \\ 
Mass scale  & - & -&$\pm 0.5$ & - \\ 
Mass resolution & - & - &- & $\pm 1.3$ \\ 
MC dipion model & $\pm 6.8$& $\pm 6.8$ & - & - \\
${\cal B}_{(\psi(2S) \to J/\psi \pipi)}$ & $\pm 1.2$& $\pm 1.2$ & - & - \\
${\cal B}_{(J/\psi \to l^{+}l^{-})}$ & $\pm 0.7$& $\pm 0.7$ & - & - \\
PID, luminosity &    & &  &  \\ 
 and tracking  &   $\pm 3.3 $ & $\pm 3.3 $  & - & - \\ 
 \hline 
Total (Y(4360))  &   $\pm 8$ &  $\pm 8$ & $\pm 9$& $\pm 13$\\ 
Total (Y(4660))  & $\pm 16$& $\pm 6$ & $\pm3$&$\pm 10$ \\ 
\end{tabular} 
\end{ruledtabular} 
\end{table} 

\begin{table} [htb!]
\caption{\label{tab:systematic4360ll} Systematic uncertainty estimates for the parameters used in the fit to data of Fig.~\ref{fig:fit}(c).} 
\begin{ruledtabular} 
\begin{tabular}{cccc} 
Source & $\Gamma_{e^+e^-}\cdot{\cal B}$  & Mass  & $\Gamma$   \\ 
 &$(\%)$  & (\mevcc) &   (\mev)  \\ 

\hline 
Fit procedure &  & &\\ 
for the Y(4360)& $\pm4$   & $\pm$3 & $\pm$ 2 \\ 
Fit procedure &  & &\\ 
for the Y(4660)& $\pm14$   & $\pm$2 & $\pm$ 3 \\ 
Mass scale  & -&$\pm 0.5$ & - \\ 
Mass resolution & - &- & $\pm 1.3$ \\ 
MC dipion model & $\pm 6.8$ & - & - \\
${\cal B}_{(\psi(2S) \to J/\psi \pipi)}$ & $\pm 1.2$ & - & - \\
${\cal B}_{(J/\psi \to l^{+}l^{-})}$ & $\pm 0.7$ & - & - \\
${\cal B}_{(\psi(2S) \to l^{+}l^{-})}$ & $\pm 1.6$ & - & - \\
PID, luminosity &   &  &  \\ 
 and tracking  &   $\pm 3.3 $   & - & - \\ 
 \hline 
Total (Y(4360))  &   $\pm 9$ &  $\pm 3$ & $\pm 3$\\ 
Total (Y(4660))  & $\pm 16$& $\pm 2$ &$\pm 4$ \\ 
\end{tabular} 
\end{ruledtabular} 
\end{table} 
Our estimates of systematic uncertainty result from the sources listed in Tables~\ref{tab:systematic4360} and \ref{tab:systematic4360ll}, where we include the latter table, which corresponds to the combined $\psi(2S)$ decay modes, for the sake of completeness.\\ 
\indent The systematic uncertainties on the fitted values of the $Y(4360)$ and the $Y(4660)$ parameters include contributions from the fitting procedure 
(evaluated by changing the fit range and the background parametrization), the uncertainty in the mass scale (which results from the uncertainties associated with the magnetic field and with our energy-loss correction procedures \cite{lambdamass,taumass}), the mass-resolution function, and the change in efficiency when the dipion distribution is simulated using the histograms in Fig.~\ref{fig:pipimass}. Uncertainties associated with luminosity, tracking, efficiency and PID affect only $\Gamma_{e^+e^-}\cdot{\cal B}$, and their net contribution is 3.3\%. Uncertainties on the relevant branching fraction values~\cite{ct:PDG} are indicated in Tables~\ref{tab:systematic4360} and \ref{tab:systematic4360ll}, and are relevant only for  $\Gamma_{e^+e^-}\cdot{\cal B}$.
These estimates of systematic uncertainty are combined in quadrature to obtain the values which we quote for the $Y(4360)$ and $Y(4660)$ states.\\
\indent In Fig. \ref{fig:pipimass} we show the $\pipi$ invariant mass distributions for events in the $\psi(2S)\pipi$, $\psi(2S)\to J/\psi \pipi$ invariant mass regions (a)~$4.0~\mathrm{GeV/c^{2}}<m_{\psi(2S)\pipi}<4.5~\mathrm{GeV/c^{2}}$, and (b) $4.5~\mathrm{GeV/c^{2}} <m_{\psi(2S)\pipi}< 4.9~\mathrm {GeV/c^{2}}$. The distributions are consistent with previous measurements~\cite{:2007ea}. In each case, the mass distribution appears to differ slightly from the phase-space expectation, as shown by the corresponding histogram. For the higher mass resonance, there is some indication of an accumulation of events in the vicinity of the $f_{0}(980)$ state. Similar behavior is observed in~\cite{:2007ea}, and both distributions bear a qualitative resemblance to the dipion invariant mass spectrum from the decay $Y(4260)\to J/\psi \pipi$ ~\cite{jpsipipi}. The small number of events involved precludes the drawing of any definite conclusion.\\ 
\indent In summary, we have used ISR events to study the reaction $e^+e^-\to\psi(2S)\pipi$ in the c.m.\ energy range 3.95--5.95 GeV. We observe two resonant structures, which we interpret as the $Y(4360)$ and the $Y(4660)$, respectively. For the $Y(4360)$ we obtain~$m=4340~\pm16~\pm~9~\mevcc$ and $\Gamma=94~\pm32~\pm~13~\mev$, and for the $Y(4660)$ $m=4669\pm21~\pm~3~\mevcc$ and $\Gamma=104~\pm48~\pm~10~\mev$; in each case the first uncertainty is statistical and the second is systematic. We thus confirm the report in Ref.~\cite{:2007ea} of a structure near 4.65 $\gevcc$, and obtain consistent parameter values for this state. If we include the $Y(4260)$, which decays to $J/\psi \pipi$~\cite{ct:babar-Y}, three charmonium-like states with $J^{PC}=1^{--}$ have been observed in the mass region 4.2-4.7~$\gevcc$, none of which has a well-understood interpretation. 
\begin{acknowledgments}
 We are grateful for the excellent luminosity and machine conditions
provided by our \pep2\ colleagues, 
and for the substantial dedicated effort from
the computing organizations that support \babar.
The collaborating institutions wish to thank 
SLAC for its support and kind hospitality. 
This work is supported by
DOE
and NSF (USA),
NSERC (Canada),
CEA and
CNRS-IN2P3
(France),
BMBF and DFG
(Germany),
INFN (Italy),
FOM (The Netherlands),
NFR (Norway),
MES (Russia),
MICIIN (Spain),
STFC (United Kingdom). 
Individuals have received support from the
Marie Curie EIF (European Union)
and the A.~P.~Sloan Foundation (USA).

\end{acknowledgments}

\bibliography{paper}

\end{document}